\documentclass[final,5p,times,twocolumn] {elsarticle}
\usepackage{lineno,hyperref}
\modulolinenumbers[5]

\journal{Journal of \LaTeX\ Templates}

\begin{document}

\begin{frontmatter}

\title{Particle Identification with DIRCs at PANDA}

\author[5]{M.~D\"{u}ren${}^*$}
\ead{michael.dueren@uni-giessen.de}
\author[1,2]{A.~Ali}
\author[1]{A.~Belias}
\author[1]{R.~Dzhygadlo}
\author[1]{A.~Gerhardt}
\author[1,2]{M.~Krebs}
\author[1]{D.~Lehmann}
\author[1,2]{K.~Peters}
\author[1]{G.~Schepers}
\author[1]{C.~Schwarz}
\author[1]{J.~Schwiening}
\author[1]{M.~Traxler}
\author[3]{L.~Schmitt}
\author[4]{M.~B\"{o}hm}
\author[4]{A.~Lehmann}
\author[4]{M.~Pfaffinger}
\author[4]{S.~Stelter}
\author[4]{F.~Uhlig}
\author[5]{E.~Etzelm\"{u}ller}
\author[5]{K.~F\"{o}hl}
\author[5]{A.~Hayrapetyan}
\author[5]{K.~Kreutzfeld}
\author[5]{J.~Rieke}
\author[5]{M.~Schmidt}
\author[5]{T.~Wasem}
\author[6]{C.~Sfienti}
\address[1]{GSI Helmholtzzentrum f\"ur Schwerionenforschung GmbH, Darmstadt, Germany}
\address[3]{FAIR, Facility for Antiproton and Ion Research in Europe, Darmstadt, Germany}
\address[4]{Friedrich Alexander-University of Erlangen-Nuremberg, Erlangen, Germany}
\address[5]{II. Physikalisches Institut, Justus Liebig-University of Giessen, Giessen, Germany}
\address[6]{Institut f\"{u}r Kernphysik, Johannes Gutenberg-University of Mainz, Mainz, Germany} 
\address[2]{Goethe Univesity Frankfurt, Frankfurt a.M., Germany}


\cortext[mycorrespondingauthor]{Corresponding author}


\begin{abstract}
The DIRC technology (Detection of Internally Reflected Cherenkov light) offers an excellent possibility to minimize the form factor of Cherenkov detectors in hermetic high energy detectors. The PANDA experiment at FAIR in Germany will combine a barrel-shaped DIRC with a disc-shaped DIRC to cover an angular range of 5 to 140 degrees. Particle identification for pions and kaons with a separation power of 3 standard deviations or more will be provided for momenta between 0.5 GeV/c and 3.5 GeV in the barrel region and up to 4 GeV/c in the forward region. 
Even though the concept is simple, the design and construction of a DIRC is challenging. High precision optics and mechanics are required to maintain the angular information of the Cherenkov photons during multiple internal reflections and to focus the individual photons onto position sensitive photon detectors. These sensors must combine high efficiencies for single photons with low dark count rates and good timing resolution at high rates. The choice of radiation hard fused silica for the optical material and of MCP-PMT photon sensors is essential for DIRC detectors to survive in an environment of radiation and strong magnetic field. The two DIRC detectors differ in the focusing optics, in the treatment of chromatic dispersion and in the electronic readout systems. 
The technical design of the two DIRC detectors and their validation by testing prototypes in a mixed particle beam at CERN are presented.
\end{abstract}

\begin{keyword}
PANDA, Cherenkov-Detector, Particle Identification
\end{keyword}

\end{frontmatter}


\section{Particle Identification of Hadrons}
The PANDA experiment~\cite{panda} at FAIR requires excellent particle identification (PID) for its rich hadronic physics program. Photons, electrons, muons and neutrons can be identified by their distinct absorption properties in the electromagnetic calorimeter and the iron yoke. Pions, kaons and protons are identified by Cherenkov detectors. The very forward region (scattering angle $\theta <5^\circ $) will be handled by a Ring Imaging Cherenkov (RICH) outside the central spectrometer. RICH detectors measure the Cherenkov rings of the individual particles to determine the velocity of the charged particles and, in combination with the momentum measurement, allow for the determination of their mass and identity. A typical RICH detector uses gas or aerogel as radiator and requires a significant amount of space behind the radiator for mirrors and/or photon detectors to allow for a precise reconstruction of the angular information of the photons. In a hermetic detector like PANDA, the PID detector has to be in between the tracking devices and the calorimeter. The significant space required by a traditional RICH leads to an increase in size and costs of the calorimeter and the magnet system. This can be avoided by using a new PID technology. 

\section{DIRC Technology}
A new concept that minimizes the transverse dimension of a Cherenkov detector to a few centimeters is the DIRC (Detection of Internally Reflected Cherenkov light) technology. A DIRC uses a solid, transparent bar or plate, typically made of fused silica, which acts as radiator and at the same time as a lightguide for Cherenkov photons. By internal total reflection, the photons are propagated to the outer rim of the radiator, where focusing elements and photon sensors are attached and allow for a determination of the photon angle outside of the sensitive area of the particle detector. 

\begin{figure*} [htb]
  \centering
    \includegraphics[width=\textwidth]{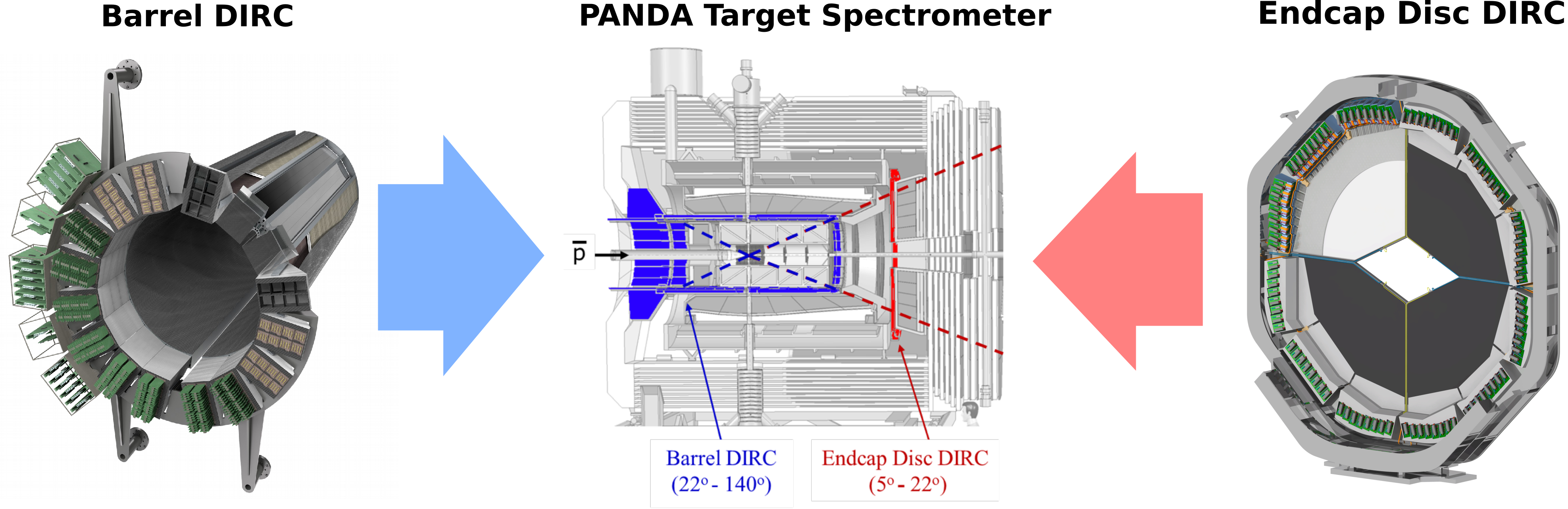}
    \caption{Particle identification in the PANDA experiment is realized in the central region by the Barrel DIRC (left) and in the forward endcap region by the Endcap Disc DIRC (right).}
\label{fig:DIRCpanda}
\end{figure*}

\begin{figure} [htb]
  \centering
    \includegraphics[width=\linewidth]{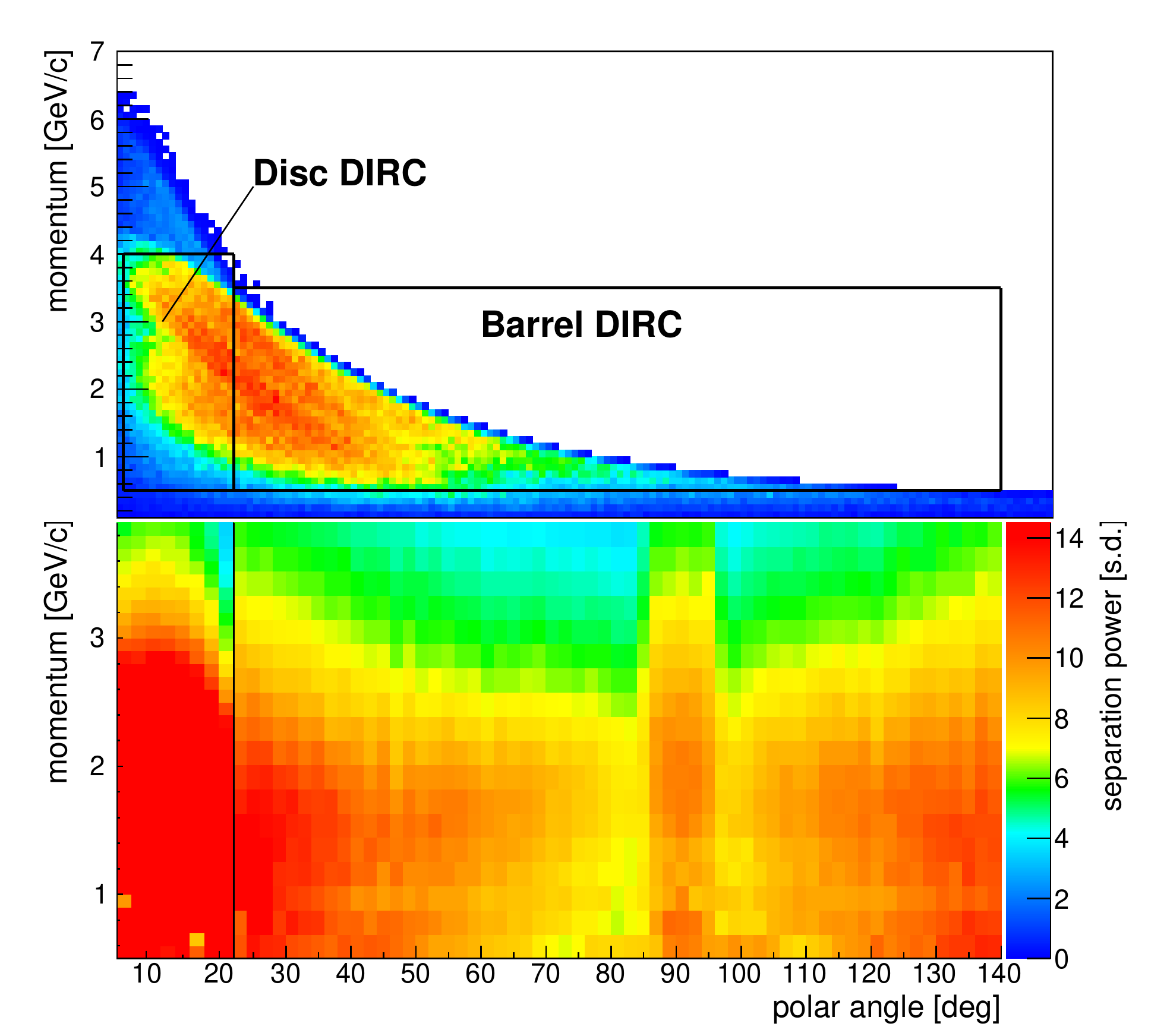}
    \caption{Upper panel: Simulated phase space distribution of kaons from eight benchmark physics channels at an antiproton momentum of $p = 7$ GeV/c. For higher (lower) beam momenta, the distribution shifts to lower (higher) polar angles across the boundary of the Barrel and the EDD DIRC coverages, which are marked with rectangles. Lower panel: Distribution of the separation power of the two DIRCs. The vertical column at $\sim 90^\circ$ results from photons that are totally reflected in the forward and in the backward direction.}
\label{fig:phasespace}
\end{figure}

PANDA will be the first experiment that implements two kinds of DIRC detectors: a Barrel DIRC (BD)~\cite{Barrel-TDR-arXiv} that covers the polar angular range of $22^\circ<\theta < 140^\circ$  and an Endcap Disc DIRC (EDD)~\cite{EDD} that covers $5^\circ <\theta <22^\circ$ (see Fig.~\ref{fig:DIRCpanda}). Fig.~\ref{fig:phasespace} (upper panel) shows that the phase space distribution of kaons in PANDA is well covered by the two DIRCs. The lower panel shows the separation power for pions and kaons. The BD (EDD) will provide a separation power of 3 standard deviations or more for momenta between 0.5 GeV/c (0.7~GeV/c) and 3.5~GeV/c (4~GeV/c). DIRC technology is well suited for high-luminosity machines, as the generation of Cherenkov photons -- in contrast to scintillation photons -- is prompt and can be detected with a precision in the order of 100 ps. DIRC detectors are challenging in three areas: optics, photon sensors and readout.

\begin{figure} [t]
  \centering
    \includegraphics[width=\linewidth]{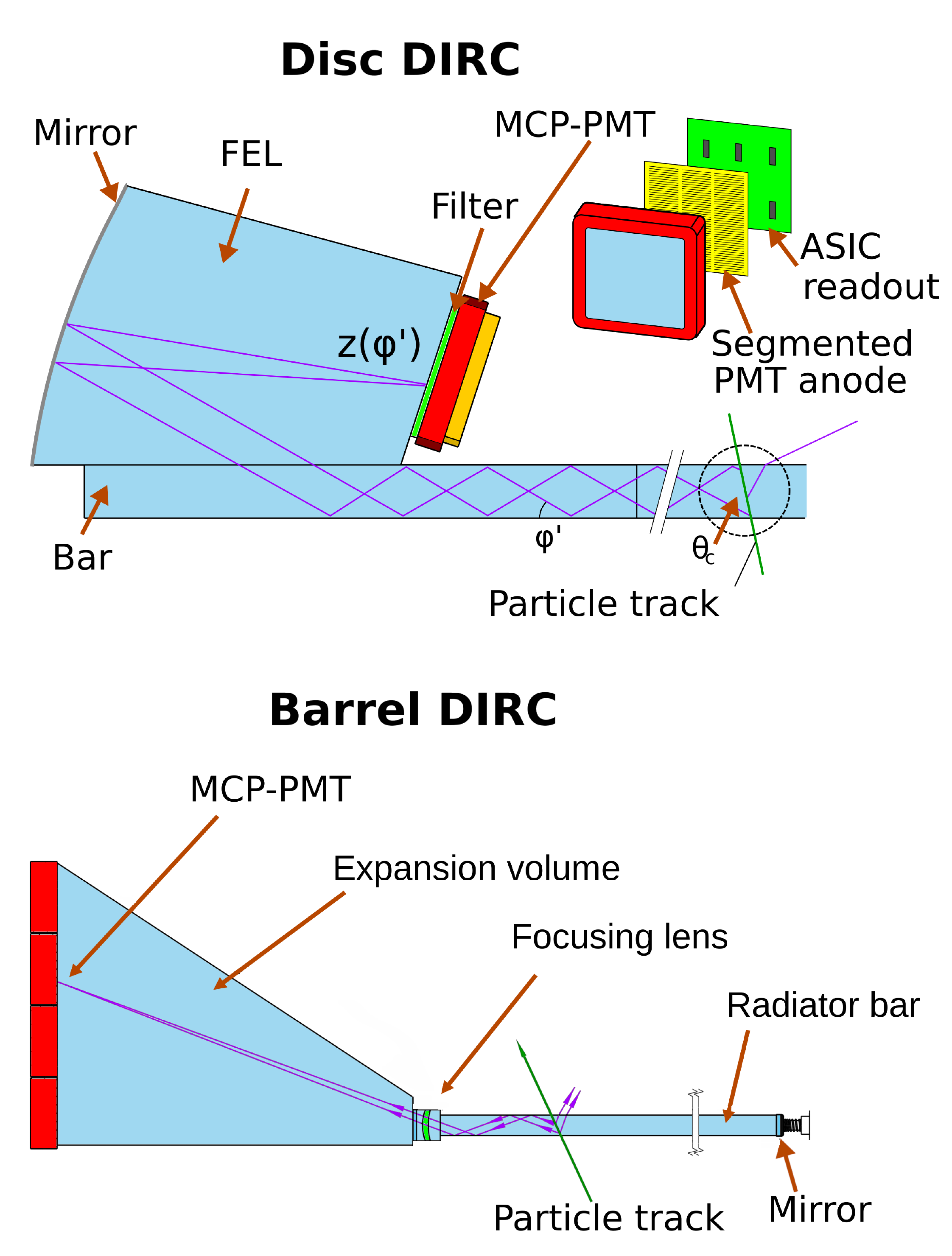}
    \caption{The charged particle produces a Cherenkov cone in the fused silica radiator. By total internal reflection, the light propagates to the region outside the acceptance of the detector. In the Endcap Disc DIRC a cylindrical mirror projects the photons onto a strip photon sensor array (upper panel). In the Barrel detector spherical focusing lenses project the photons via an expansion prism onto a photon sensor matrix (lower panel).}
\label{fig:DIRCoptics}
\end{figure}

\paragraph{Optics}  The Cherenkov light is produced when the charged particles cross the 1.7~cm (2~cm) thick radiators of the BD (EDD). High precision optics and mechanics are required to maintain the angular information of the Cherenkov photons during multiple internal reflections. The Cherenkov radiator in the BD consists of 240~cm long bars which are coupled to multi-layer spherical focusing lenses and prisms to project the individual photons onto position sensitive photon sensors (see Fig.~\ref{fig:DIRCoptics}, lower panel). The EDD uses a large plate with an outer diameter of 210~cm in the shape of a dodecagon as radiator. To simplify the production of the large 2~cm thin plate, the EDD is divided into 4 independent identical quadrants that are optically decoupled. Focusing is done by cylindrical, aluminized mirror elements (see Fig.~\ref{fig:DIRCoptics}, upper panel). All optical elements (except the lenses of the BD) are made from radiation hard synthetic fused silica with excellent surface roughness of $<5$ \r{A}  rms. The focusing lens of the BD is a triplet made from one layer of lanthanum crown glass (NLaK33, refractive index $n$=1.786 for $\lambda$=380 nm) between two layers of synthetic fused silica ($n$=1.473 for $\lambda$=380 nm). By avoiding air gaps and the use of a large-$n$ material, photon losses at the surfaces are minimized.

\paragraph{Photon sensors} The photon sensors must combine high efficiencies for single photons with low dark count rates and good timing resolution at high rates. Multi anode MCP-PMT photon sensors are currently the only choice that fulfil all requirements concerning radiation hardness, rates, lifetime and the strong magnetic field~\cite{lehmann2016}. The BD (EDD) will use a spatial segmentation of $8\times 8$ (approx. $3\times 100$) pixel and a pitch of 6.2 (0.5) mm. In the EDD, optical filters and/or special photocathodes will be used to limit the spectral range of the sensors in order to reduce chromatic smearing. There is a trade-off concerning single photon resolution and photon statistics.

\paragraph{Readout} The readout electronics of the two DIRCs are trigger-less, free running systems that register and identify multi-photon coincidences as candidates for particles that produce Cherenkov light. The digitization for the BD is realized by a TDC-in-FPGA technology (DiRICH)~\cite{trb3,dirich}, while the EDD uses the TofPET2 ASIC~\cite{tofpet}, that allows for a highly integrated compact backplane. The reconstruction of the Cherenkov angle from the measured hit pattern requires sophisticated algorithms. Several methods have been developed using track-by-track likelihood tests based on the position and the arrival time of the photons~\cite{roman, mustafa_jinst}.  In the real experiment, a first stage real-time reconstruction will be needed to select interesting physics events from a large hadronic background before data are written to disc.

\section{Testbeam Results}

Several prototypes of the BD and EDD have been produced and tested in mixed hadronic particle beams at CERN and electron beams at DESY. The latest campaign at the T9 beamline at CERN PS in August 2018 was a direct measurement of the PID performance of BD and EDD across the PANDA phase space. The particles of a mixed pion-proton beam with variable momentum were identified using time-of-flight counters.

\begin{figure} [t]
  \centering
    \includegraphics[width=\linewidth]{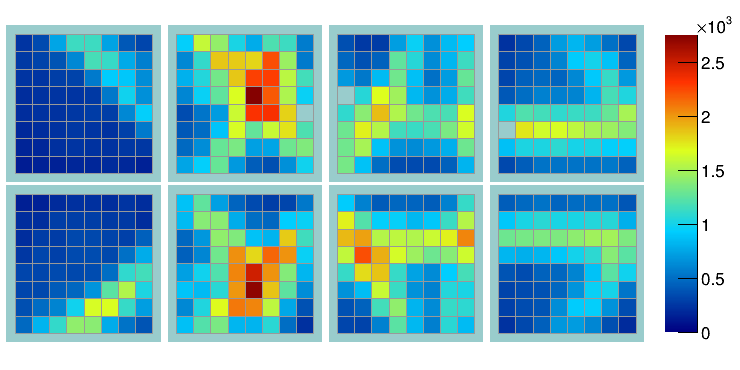}
    \caption{The measured barrel DIRC photon patterns of the 2018 test beam data are shown. By overlaying many events, a 7~GeV/c pion beam on a barrel DIRC prototype generates a complicated pattern which results from the multiple reflections of the Cherenkov light during the propagation in the $17 \times 53 $ mm$^2$ radiator bars.}
\label{fig:DIRCpattern1}
\end{figure}

\begin{figure} [t]
  \centering
   \includegraphics[width=\linewidth]{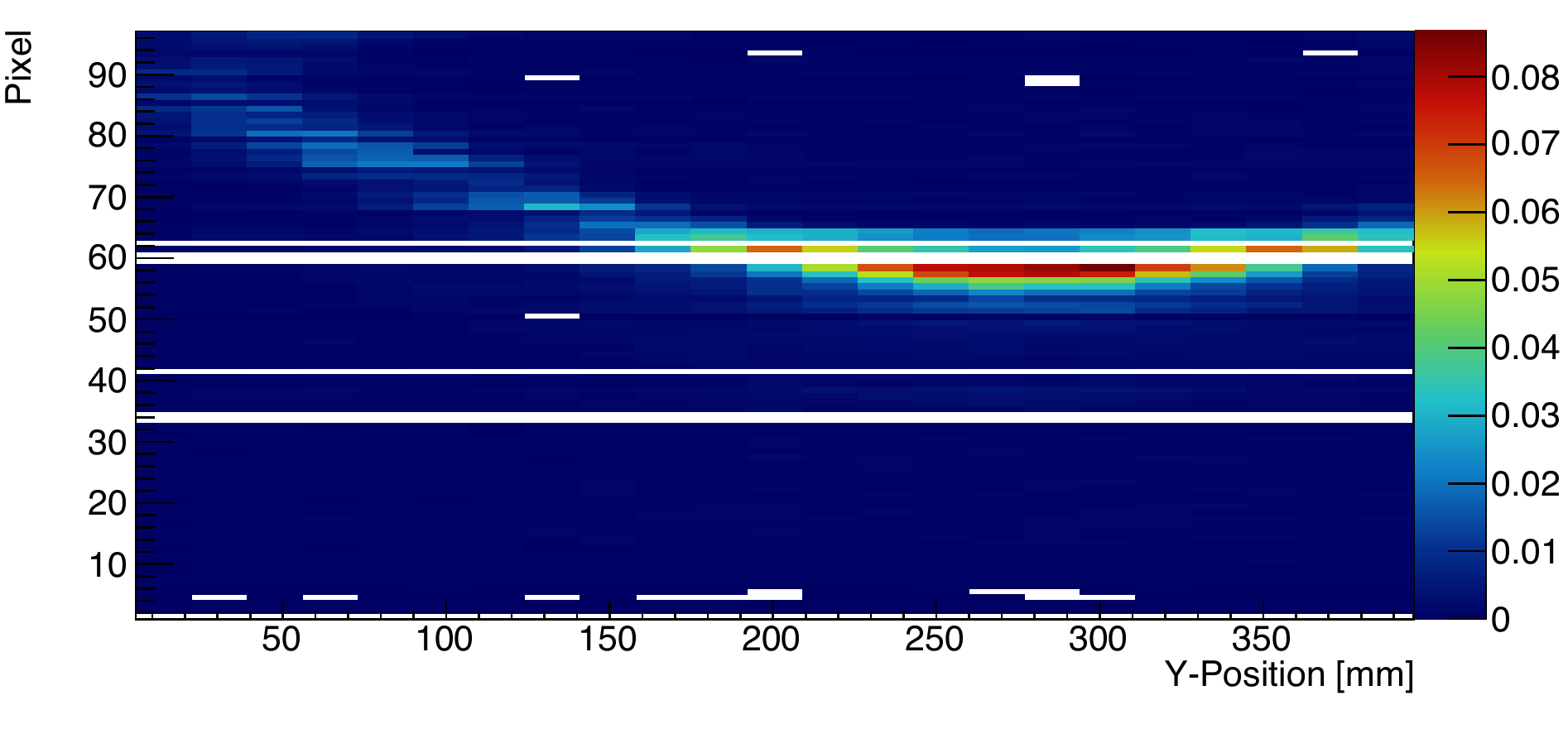}
    \caption{The measured Endcap Disc DIRC photon patterns of the 2018 test beam data are shown. A vertical scan of the 10\,GeV/c pion/proton beam on a $500\times 500\times 20$ mm$^3$ rectangular radiator plate generates a pattern in one MCP-PMT and emulates the pattern of an EDD detector that is fully equipped with MCP-PMTs.}
\label{fig:DIRCpattern2}
\end{figure}

\paragraph{DIRC pattern} Figure~\ref{fig:DIRCpattern1} shows the pattern of the Cherenkov photons in the BD as seen by its MCP-PMT matrix. The complicated pattern comes from multiple reflections at the lateral faces of the radiator bars. For cost saving reasons, the EDD prototype used a small ($50\times 50$ cm$^2$) quadratic radiator plate which was only partially equipped with focusing elements. Instead of measuring the photon distribution on all sensor positions for a fixed beam position, the beam was scanned across the plate (along the vertical y-axis) and measured by only one sensor at a fixed position with regard to the radiator plate. 
Figure~\ref{fig:DIRCpattern2} shows the pixel position of the Cherenkov photons as a function of the beam position. The minimum of the `smile' is at the position where the beam and the sensor have the same y-coordinate. 

\begin{figure} [t]
  \centering
    \includegraphics[width=\linewidth]{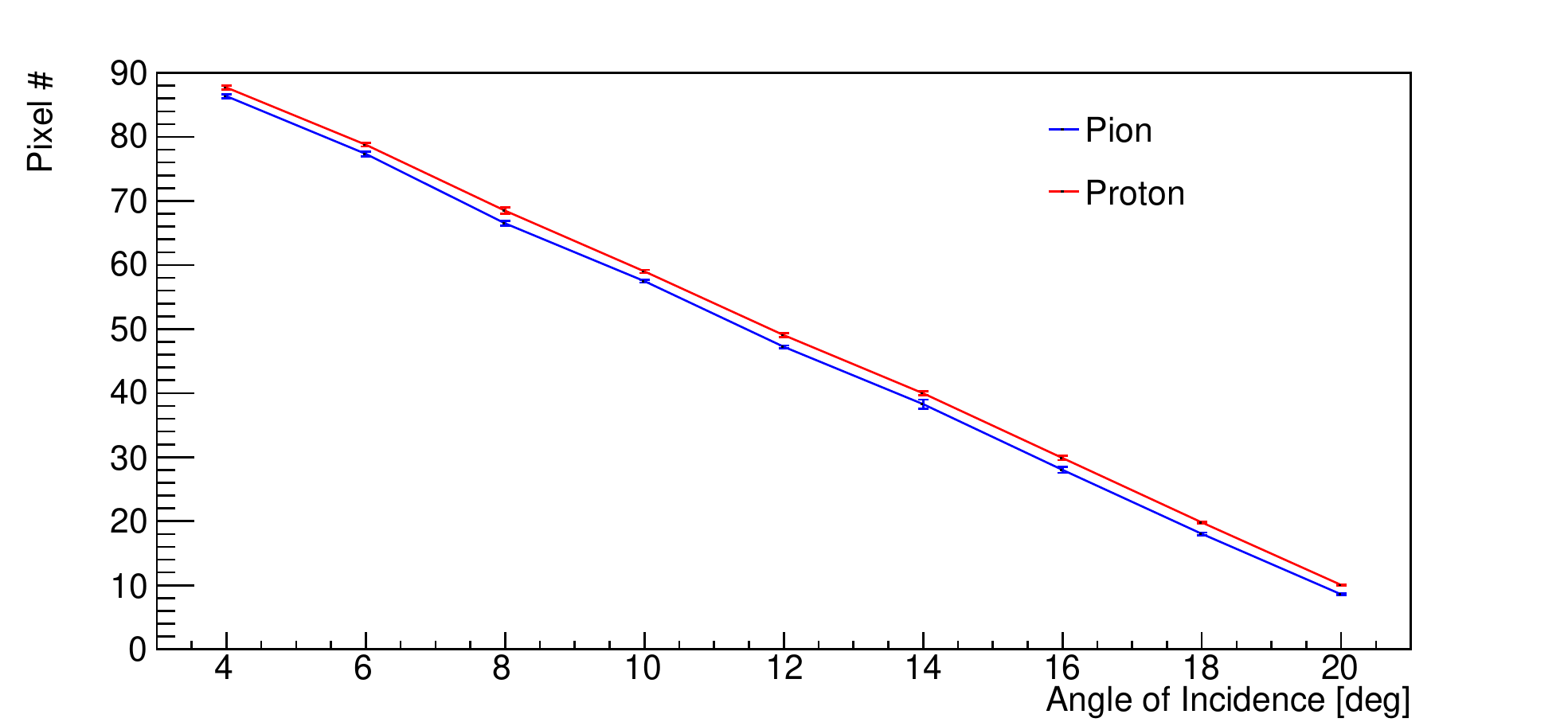}
    \caption{Preliminary results from the 2018 test beam campaign of the EDD prototype. The measured pixel position on the MCP-PMT is plotted as a function of the beam angle. The beam momentum was set to 7\,GeV/c. TOF counters have been used to discriminate pions and protons.}
\label{fig:ilknur1}
\end{figure}

\begin{figure} [t]
  \centering
    \includegraphics[width=\linewidth]{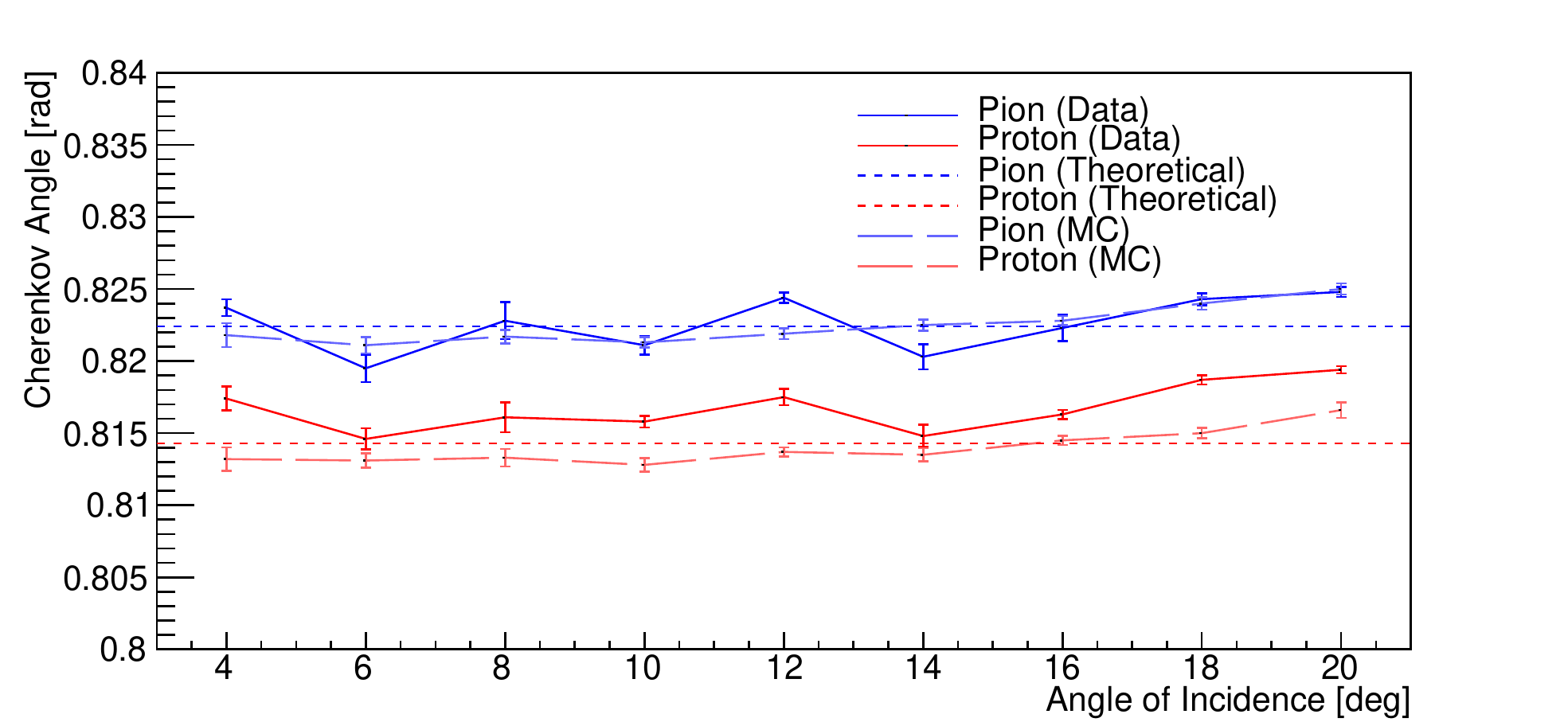}
    \caption{Preliminary results from the 2018 test beam campaign of the EDD prototype. The reconstructed Cherenkov angle for pions and protons. The data agree roughly with the Monte Carlo simulation and with the theoretical values for pions and protons at 7\,GeV/c. The deviations between the proton data and the predictions are under investigation.}
\label{fig:ilknur2}
\end{figure}

\paragraph{Endcap Disc DIRC} By rotating the EDD radiator plate around its vertical axis, the pixel number changes with the angle of incidence. Figure~\ref{fig:ilknur1} shows the linear relation between the pixel number and the particle angle, as well as a small shift of the pattern for pions compared to protons at a momentum of 7 GeV/c. From these values, the reconstruction software extracts the Cherenkov angle by taking all accessible geometric effects in the radiator and the lightguide into account as shown in Figure~\ref{fig:ilknur2}. As expected, the reconstructed pion data at 7 GeV/c agree with the Monte Carlo results and with the theoretical Cherenkov angles. Small deviations come from statistical effects and from the finite segmentation of the lightguides. The systematic shift of the proton data by $\sim2$~mrad is under investigation and may be related to an insufficient pion-proton separation in the TOF analysis of the EDD runs. 

\paragraph{Barrel DIRC}

\begin{figure} [thb]
  \centering
    \includegraphics[width=\linewidth]{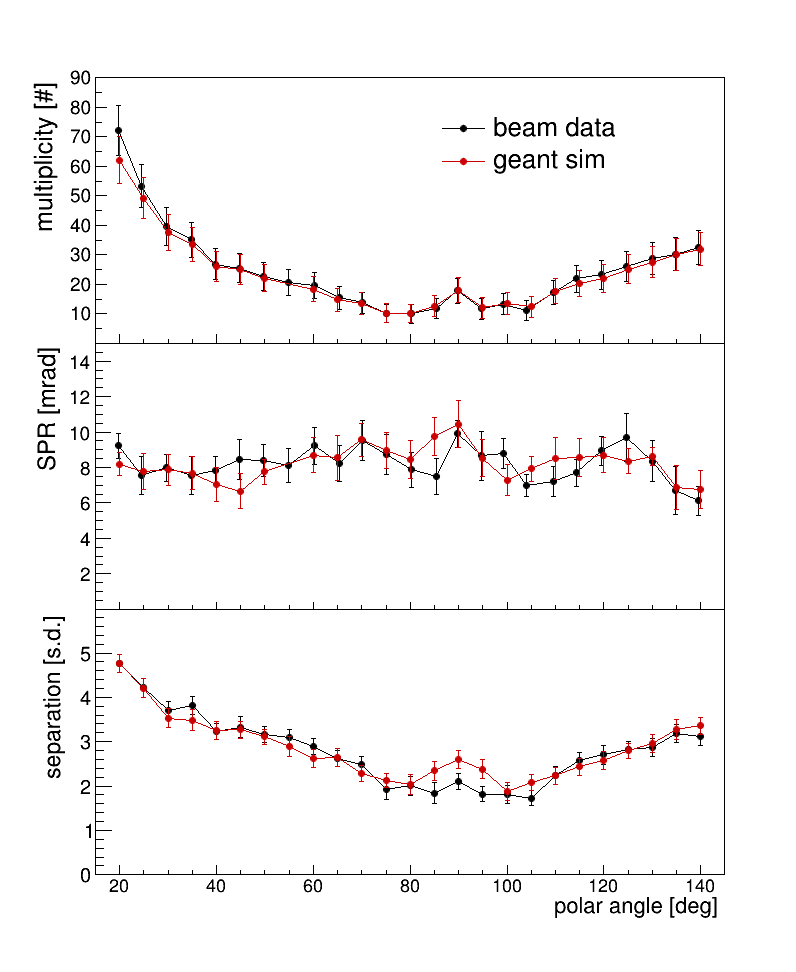}
    \caption{The angular scan of a 7 GeV/c mixed pion-proton beam crossing the barrel DIRC prototype. Shown are the photon yields (upper panel), the single photon resolution (middle), and the separation power between pions and protons (lower panel) as a function of the incident hadron angle at the radiator bar. The error bars are dominated by systematic uncertainties. There is good agreement of the 2018 test beam data with the corresponding Geant4 simulations.}
\label{fig:BDresults}
\end{figure}

The aim of the BD measurements was to validate the properties of its near final design. 
Figure~\ref{fig:BDresults} shows the multiplicity of the Cherenkov photons (top panel), the angular resolution of the single photon measurement (middle panel), and the separation power of the pion-proton identification as a function of the incident particle angle. There is excellent agreement between the measured results and the performance simulated with a Geant4 Monte Carlo\cite{geant}. As the Cherenkov angle separation of pion-proton at 7~GeV/c corresponds to a kaon-pion-separation at about 3.5~GeV/c, the successful results from the  mixed pion-proton testbeam can be directly used to assess the final performance of the DIRCs in PANDA.

\section*{Acknowledgements}
This work was funded by BMBF and HIC for FAIR. We would also like to thank the DESY and CERN staff for the opportunity to use the local test beam facilities and their on-site support as well as the VCI2019 organizers in Vienna.


\end{document}